\documentclass[preprint,12pt, a4paper]{elsarticle}

\usepackage{amssymb}
\usepackage[colorlinks=true, urlcolor=blue]{hyperref}
\usepackage{amsthm}
\usepackage{threeparttable}

\usepackage{xurl}
\usepackage{seqsplit}

\usepackage{float}
\usepackage{multirow}
\usepackage{pifont}
\usepackage{graphicx}
\usepackage{listings}
\usepackage[table,xcdraw]{xcolor}
\newcommand{\cmark}{\ding{51}}
\newcommand{\xmark}{\ding{55}}

\definecolor{codegreen}{rgb}{0,0.6,0}
\definecolor{codegray}{rgb}{0.5,0.5,0.5}
\definecolor{codepurple}{rgb}{0.58,0,0.82}
\definecolor{backcolour}{rgb}{0.95,0.95,0.92}

\lstdefinestyle{mystyle}{
    backgroundcolor=\color{backcolour},   
    commentstyle=\color{codegreen},
    keywordstyle=\color{magenta},
    numberstyle=\tiny\color{codegray},
    stringstyle=\color{codepurple},
    basicstyle=\ttfamily\footnotesize,
    breakatwhitespace=false,         
    breaklines=true,                 
    captionpos=b,                    
    keepspaces=true,                 
    numbers=left,                    
    numbersep=5pt,                  
    showspaces=false,                
    showstringspaces=false,
    showtabs=false,                  
    tabsize=2
}

\usepackage{lineno}

\journal{SoftwareX}

\begin{document}
\renewcommand{\labelenumii}{\arabic{enumi}.\arabic{enumii}}

\begin{frontmatter}

\title{\texttt{SurVigilance}: An Application for Accessing Global Pharmacovigilance Data}

\author[1,2]{Raktim Mukhopadhyay\corref{corresponding}}
\ead{raktimmu@buffalo.edu}
\author[1,3]{Marianthi Markatou}
\ead{markatou@buffalo.edu}
\address[1]{Department of Biostatistics, University at Buffalo, NY, USA}
\address[2]{Institute for Artificial Intelligence and Data Science, University at Buffalo, NY, USA}
\cortext[corresponding]{Corresponding author.}
\address[3]{Department of Medicine, Jacobs School of Medicine and Biomedical Sciences, University at Buffalo, NY, USA}

\begin{abstract}
Even though several publicly accessible pharmacovigilance databases are available, extracting data from them is a technically challenging process. Existing tools typically focus on a single database. We present \texttt{SurVigilance}, an open-source tool that streamlines the process of retrieving safety data from seven major pharmacovigilance databases. \texttt{SurVigilance} provides a graphical user interface as well as functions for programmatic access, thus enabling integration into existing research workflows. \texttt{SurVigilance} utilizes a modular architecture to provide access to the heterogeneous sources. By reducing the technical barriers to accessing safety data, \texttt{SurVigilance} aims to facilitate pharmacovigilance research. 
\end{abstract}


\end{frontmatter}



\section{Introduction}

According to the World Health Organization (WHO), “first, do no harm” is one of the most fundamental principles of healthcare \citep{Patients39:online}. However, individuals experience adverse events (AEs) after administration of drugs, vaccines, and other therapeutic biologics. This raises a serious public health concern, which is associated with increased costs \citep{white1999counting}, and results in hospitalizations and deaths \citep{Preventa41:online}. Clinical trials provide the highest standard of evidence for safety information on drugs, vaccines, and other therapeutic biologics before a drug is licensed by regulatory authorities. However, clinical trials suffer from a number of limitations that affect their generalizability \citep{Amery1999-gx}, and hence underscore the need for post-marketing surveillance. \\

The identification of safety issues related to a product after its introduction to the market is primarily conducted using the data obtained from spontaneous reporting systems (SRS), examples of which include the Food and Drug Administration (FDA) Adverse Event Reporting System (FAERS) and VigiBase, maintained by the WHO, among others. SRS are passive surveillance systems that collect information on potential safety issues associated with the use of medical products submitted by healthcare workers, patients and their caregivers. For several decades, pharmaceutical corporations, public health, and regulatory authorities have relied heavily on SRS databases to conduct post-marketing surveillance of medications, vaccines, biologics and medical devices. \\

Several methods of disproportionality analysis have been developed and applied primarily to data derived from such safety databases. These methods include Proportional Reporting Ratio, Reporting Odds Ratio, Bayesian Confidence Propagation Neural Network, Multi-item Gamma Poisson Shrinker, Likelihood Ratio Test (LRT)-based methods, to name a few. A fairly complete description of the methods appears in \citep{Ding2020-ul}, while various LRT-based methods can be found in \citep{Huang2014-zl, Saptarshi_Chakraborty_Anran_Liu_Robert_Ball_Marianthi_Markatou2022-ek}, with TreeScan \citep{treescan} being an alternative method. Regardless of the data representation, all of these methods require count data associated with the suspected AEs in order to be applied. \\

{Let's assume that a contingency table contains $I$ suspected AEs ($i = 1, 2, \ldots, I$) associated with $J$ drugs ($j = 1, 2, \ldots, J$). An illustration of the $2 \times 2$ contingency table used in these methods is shown in Table \ref{tab:ae-contingency}. The count of the $i$-th AE and $j$-th drug combination is represented by $n_{ij}$. The marginal row total of the $i$-th AE is represented as $n_{i \cdot}$, while the column marginal of the $j$-th drug is $n_{\cdot j}$. Mathematically, $n_{i \cdot} = \sum_{j=1}^{J} n_{ij}$ and $n_{\cdot j} = \sum_{i=1}^{I} n_{ij}$. The total number of suspected AEs in the contingency table is $n_{\cdot \cdot}$. The authors in \citep{Ding2020-ul} introduced two different methods of contingency table construction, namely AE-based and drug-based comparisons. A brief discussion on this aspect is included in Section \ref{sec:construction-of-contingency-table}.}

\begin{table}[H]
\centering
\begin{tabular}{|>{\columncolor[HTML]{C0C0C0}}l|c|c|c|}
\hline
\rowcolor[HTML]{C0C0C0}
 & \textbf{$Drug_j$} & \textbf{Other Drugs} & \textbf{AE marginal total} \\ \hline
$AE_i$              & $n_{ij}$ & $n_{i \cdot} - n_{ij}$ & $n_{i \cdot}$ \\ \hline
Other AEs           & $n_{\cdot j}-n_{ij}$ & $n_{\cdot \cdot} - n_{i \cdot} - n_{\cdot j} + n_{ij}$ & $n_{\cdot \cdot} - n_{i \cdot}$ \\ \hline
Drug marginal total & $n_{\cdot j}$ & $n_{\cdot \cdot} - n_{\cdot j}$ & $n_{\cdot \cdot}$ \\ \hline
\end{tabular}
\caption{Contingency table for AE-based signal detection}
\label{tab:ae-contingency}
\end{table}

The accessibility of eleven open-access pharmacovigilance databases was investigated in 2016 \citep{Fouretier2016}. {Out of the eleven databases investigated in the study, our tool, \texttt{SurVigilance}, currently supports the seven databases described below:}

\begin{itemize}
    \item {\textbf{Australian Database of Adverse Event Notifications (DAEN):} DAEN (medicines) contains AE data associated with medicines or vaccines. The database provides both aggregated data and detailed reports organized chronologically for a specific medicine or vaccine. The DAEN (medicines) has data dating back to January 1971, while the DAEN (medical products) includes reports starting from July 1, 2012 \citep{Database45:online}. We currently support access to data from the DAEN (medicines) database in the tool. The database can be accessed at \url{https://www.tga.gov.au/safety/database-adverse-event-notifications-daen}.}. 

    \item {\textbf{Danish Medicines Agency (Lægemiddelstyrelsen) Database \\(DMA):} DMA provides aggregated data on suspected AEs reported in Denmark for drugs. The data is derived from the DMA pharmacovigilance database, which contains reports since 1968 \citep{dma}. The database can be accessed at \url{https://laegemiddelstyrelsen.dk/en/sideeffects/side-effects-of-medicines/interactive-adverse-drug-reaction-overviews/}.}

    \item \textbf{\textbf{Netherlands Lareb Database}}: The Netherlands pharmacovigilance centre Lareb maintains a database of AEs for both drugs and vaccines. The database was established in 1995; since then, it has received close to 20,000–25,000 reports annually from patients, healthcare professionals, and pharmaceutical companies \citep{netherland-organization, TheNethe30:online}. The database can be accessed at \url{https://www.lareb.nl/en/}.

    \item {\textbf{New Zealand Medsafe Database:} The Medsafe database contains spontaneous reports submitted in New Zealand since 1965 \citep{Fouretier2016}. The database provides access to both aggregated data and de-identified spontaneous reports for a drug or vaccine of interest. The database can be accessed at \url{https://www.medsafe.govt.nz/SMARS/Default}.}

    \item \textbf{\textbf{United States Food and Drug Administration (FDA) Adverse Event Reporting System (FAERS)}}: The FAERS database contains the spontaneous reports submitted to the FDA. FAERS provides access to the de-identified spontaneous reports submitted by patients, healthcare providers and pharmaceutical companies \citep{FDAsAdve38:online}. Quarterly updates of the database can be found at \url{https://fis.fda.gov/extensions/FPD-QDE-FAERS/FPD-QDE-FAERS.html}.

    \item \textbf{\textbf{United States Vaccine Adverse Event Reporting System \\(VAERS)}}: VAERS contains reports of AEs following immunization. This database is maintained by the Centers for Disease Control and Prevention (CDC) and FDA \citep{vaers}. The VAERS database can be accessed at \url{https://vaers.hhs.gov/data/datasets.html}. 

    \item \textbf{\textbf{WHO VigiAccess}}: VigiBase, maintained by the WHO, is the largest pharmacovigilance database in the world, with $>$30 million reports submitted since 1968 by member countries of the WHO Programme for International Drug Monitoring (PIDM) \citep{VigiBase85:online}. In 2015, the WHO launched VigiAccess to provide public access to aggregated data from VigiBase \citep{Shankar2016-ta}.  The database can be accessed at \url{https://www.vigiaccess.org/}.
\end{itemize}

{The authors in \citep{Fouretier2016} found that databases in the United States offered a high level of access, while the databases from New Zealand and Australia provided a medium level of access. The databases maintained by the WHO, the Netherlands, and Denmark provided comparatively limited access. United States databases such as FAERS and VAERS grant access to individual case safety reports (ICSRs) as well as allow full database downloads, which also facilitates the calculation of population denominators and represents a high degree of transparency. Australia and New Zealand provide access to certain demographic information such as age and drugs/vaccines contained in each ICSR, and hence are considered to provide a medium level of access. In contrast, the databases from the WHO, the Netherlands, and Denmark only provide aggregated data on AE counts per drug and summarized demographic information. Since these databases do not provide complete database downloads and access to all ICSRs, computing the population denominator is more complex. In addition to the databases supported by \texttt{SurVigilance}, other publicly available databases are European Medicines Agency (EMA) EudraVigilance, United Kingdom (UK) Yellow Card, and Canadian MedEffect databases. The EudraVigilance database uses a legacy system to provide access to data, while the UK database presents the data in a highly complex format. As a result, accessing data from these sources is significantly more challenging than others and hence is not included in \texttt{SurVigilance}. The MedEffect database is available as a single download from \url{https://www.canada.ca/en/health-canada/services/drugs-health-products/medeffect-canada/adverse-reaction-database/canada-vigilance-online-database-data-extract.html}, and hence has not been included in the tool due to the simplicity in accessing the data.}\\

\begin{table}[!h]
\begin{tabular}{|l|p{6.5cm}|p{6.5cm}|}
\hline
\textbf{Nr.} & \textbf{Code metadata description} & \textbf{Metadata} \\
\hline
C1 & Current code version & 1.0.0 \\
\hline
C2 & Permanent link to code/repository used for this code version & \url{https://github.com/rmj3197/SurVigilance} \\
\hline
C3  & Permanent link to Reproducible Capsule & \url{https://codeocean.com/capsule/7678655/tree/v1}\\
\hline
C4 & Legal Code License   & GNU General Public License (GPL-3.0) \\
\hline
C5 & Code versioning system used & git \\
\hline
C6 & Software code languages, tools, and services used & \texttt{Python}, \texttt{Streamlit}, \texttt{SeleniumBase} \\
\hline
C7 & Compilation requirements, operating environments \& dependencies & Supports multiple operating systems; Needs installed Google Chrome web browser; Internet access is required for application to access the databases; \texttt{Python>=3.10, <=3.13}, \texttt{Python} packages: \texttt{beautifulsoup4}, \texttt{lxml}, \texttt{openpyxl}, \texttt{pandas}, \texttt{requests}, \texttt{seleniumbase}, \texttt{streamlit}\\
\hline
C8 & If available, link to developer documentation/manual & \url{https://survigilance.netlify.app/}\\
\hline
C9 & Support email for questions & \texttt{raktimmu@buffalo.edu}\\
\hline
\end{tabular}
\caption{Code metadata}
\label{codeMetadata} 
\end{table}

{In this paper, we present \texttt{SurVigilance}, a tool that allows users to access data from multiple safety databases across the world. The primary goal of this tool is to facilitate pharmacovigilance researchers’ access to safety data. Many existing safety databases restrict users by allowing them only to view and not download the data (e.g., WHO VigiAccess). \texttt{SurVigilance} addresses this limitation by enabling researchers to download safety data. \texttt{SurVigilance} reduces the complexity of navigating different databases while streamlining the process of data retrieval for further analysis. The tool supports two complementary modes of use: first, a simple and user-friendly graphical user interface that removes the barrier for non-technical users; and second, a set of functions that provide programmatic access to the same data retrieval functionalities. The multiple access methods allow \texttt{SurVigilance} to be integrated into workflows and larger data pipelines, making it suitable for both exploratory analysis and reproducible research. We hope that this tool aids the scientific community in the investigation of safety concerns associated with various medical products.}



\section{Software description}

\begin{figure}[!ht]
    \centering
    \includegraphics[scale=0.37]{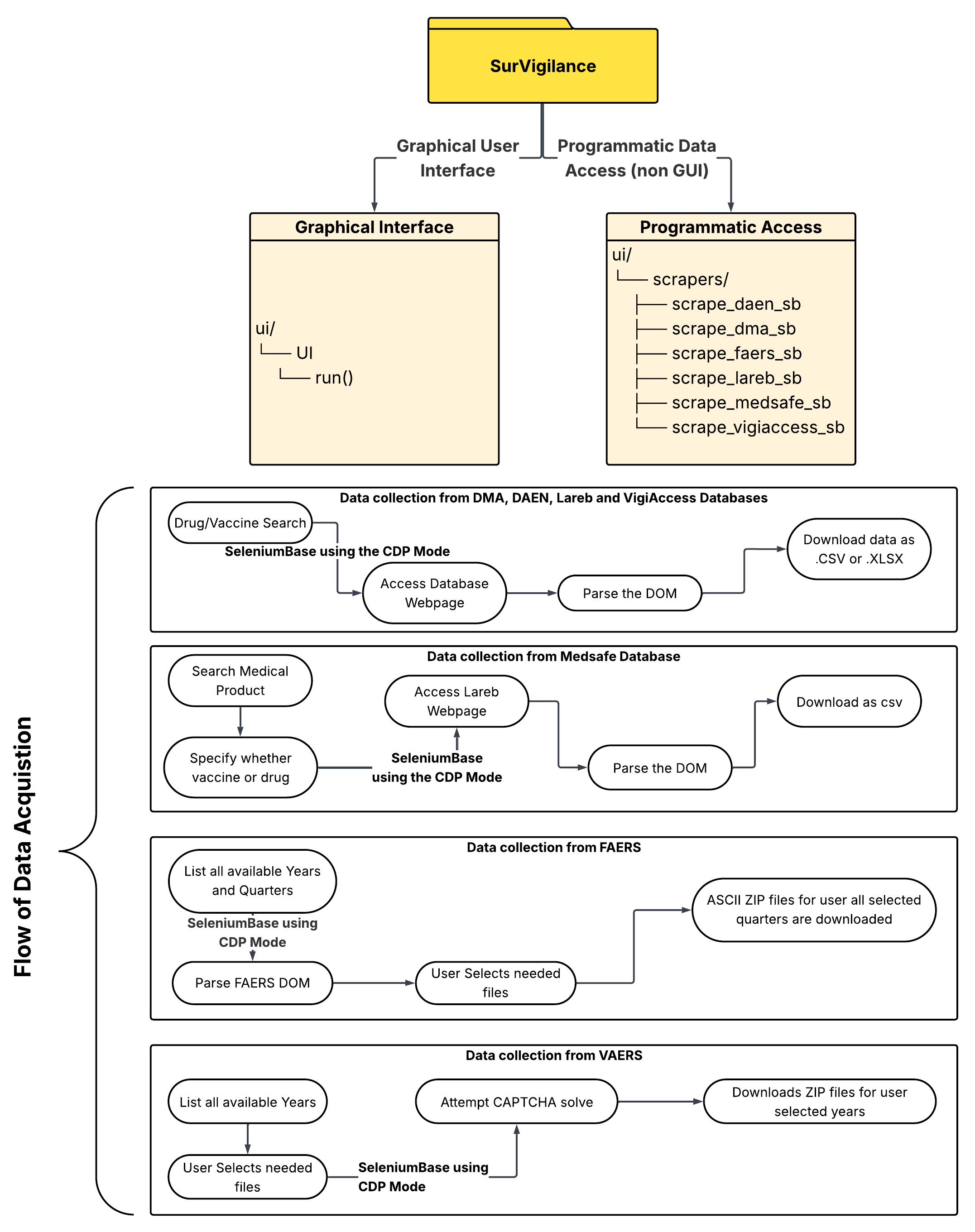}
    \caption{A simplified overview of the main functionalities of \texttt{SurVigilance} is depicted. \texttt{SurVigilance} provides the graphical interface using the \texttt{UI} class of the \texttt{ui} module. The scraping functions implemented in \texttt{scrapers} sub-module of the \texttt{ui} module can also be accessed separately to perform data collection in a programmatic manner. Additionally, the process of data acquisition from each database is illustrated. The abbreviation DOM stands for Document Object Model, which represents the structure of a webpage.}
    \label{fig:SurVigilance-schematic}
\end{figure}

\texttt{SurVigilance} is implemented using the \texttt{Streamlit} framework in the \texttt{Python} programming language. Since \texttt{SurVigilance} is implemented using \texttt{Streamlit}, the application needs only a web browser to run, removing complications associated with additional installers and hardware-dependent customizations (e.g., \texttt{arm64}- or \texttt{x86\_64}-specific installers). \texttt{SurVigilance} provides access to seven databases: DAEN, DMA, Lareb, Medsafe, FAERS, VAERS, and VigiAccess. In order to provide access to data from these databases, \texttt{SurVigilan\-ce} uses a modular approach, where the code for each safety database is kept independent of the others. This enhances the readability of the code and ensures that the codebase is maintainable and extensible for future development. A schematic diagram highlighting the various functionalities and the process of accessing the various databases is depicted in Figure \ref{fig:SurVigilance-schematic}. Salient features of the software are highlighted below: 

\begin{itemize}
    \item \textbf{User Interface:} As noted previously, an easy-to-use user interface is created using the \texttt{Streamlit} framework. The \texttt{UI} class of the \texttt{ui} module provides the graphical interface as shown in Figure \ref{fig:SurVigilance-schematic}. The \texttt{\_app.py} file provides the home page of the application. The home page provides users the options to select the location where data files will be saved and to select a database from which data will be accessed. Upon selecting the relevant database, the user is then taken to the respective page dedicated to the database. Each database has a dedicated page located under \texttt{ui/pages} that provides the graphical interface. These pages include the relevant calls to functions and customizations needed for downloading the data from the different safety databases. For the DAEN, DMA, Medsafe, Lareb, and VigiAccess databases, we provide the option to search for specific drugs or vaccines for which data needs to be downloaded. In contrast, for the FAERS and VAERS databases, we list all the years for which the data is available and download the entire database for that period. This difference stems from the manner in which each safety database permits access to its underlying data.

    \item \textbf{Scrapers and Programmatic Access:}  The code that enables \texttt{SurVig\-ilance} to scrape or download the data lives in the \texttt{scrapers/} submodule within \texttt{ui/}, depicted in Figure \ref{fig:SurVigilance-schematic}. The various functions in this submodule enable programmatic access to the data retrieval methods without the need for graphical interfaces. An overview of how the various scrapers work is also included in Figure \ref{fig:SurVigilance-schematic}. A discussion on the web-automation/scraping is included in Section \ref{subsec:scraping}. 

    \item \textbf{Data Storage:} The application allows users to specify a location where the downloaded data will persist even after the application is closed. A sub-folder for each database is created inside the location, in which the downloaded data is stored. The data from \texttt{FAERS} and \texttt{VAERS} databases are stored in a compressed file format (\texttt{.zip}), as they include export of the entire database comprising of multiple files, and made available in that format on the \texttt{FAERS} and \texttt{VAERS} webpages. The data from DAEN is stored as \texttt{.xlsx} format as the DAEN webpage exports the data in that format. Data from all other databases are stored as \texttt{.csv} after extracting the data from their respective webpages. The other databases do allow to create exports of the data or download the data.
    
\end{itemize}

\subsection{Web-automation/Scraping}\label{subsec:scraping}

The web-automation, scraping and downloading functionality of the tool is built using \texttt{SeleniumBase} \citep{Selenium4:online}. \texttt{SeleniumBase} is a browser automation framework built on top of \texttt{Selenium}. Our choice to use \texttt{SeleniumBase} is associated with a number of advantages. {Browser automation generally requires a WebDriver\footnote{WebDrivers are open-source tools that provide a programmatic interface to interact with a browser in order to accomplish various web-automation tasks. Details can be found in \citep{WebDrive43:online}.} that is compatible with the installed web browser.} \texttt{SeleniumBase} manages its own drivers, thereby eliminating the complexity for users of providing access to an appropriate WebDriver. {\texttt{SeleniumBase} offers two different ways of interacting with the browser, namely the \texttt{undetected\-chromedriver} (UC) and \texttt{Chrome DevTools Proto\-col} (\texttt{CDP}) modes \footnote{Please see \citep{Alookbac32:online} for a detailed explanation of WebDriver and \texttt{CDP}.}. These options allow robust web automation needed for retrieving the data. Please see \ref{app:appendix-responsible} for a discussion on the various measures employed for responsible web scraping.}\\

{The various webpages associated with the databases allow access to data only after performing a set of actions in a specific sequence, and hence a primary challenge was handling interactions with various web elements present on a webpage. To ensure reproducible data extraction, we intentionally added wait times at various stages of the scraping process. This ensures that each required web element is fully loaded and behaves as expected before we attempt to interact with it in order to retrieve the data. While FAERS and VAERS provide access to the data files through dedicated endpoints and can be downloaded simply using the \texttt{requests} module, the other databases rely on dynamically rendered web components that appear only after certain actions are performed. The extensive use of dynamic web elements complicates the extraction of data from the Document Object Model (DOM). This necessitated careful analysis of the structure of each webpage using browser developer tools, specifically \texttt{Chrome DevTools}. We also note that, for the VAERS database, users must complete a verification CAPTCHA when downloading data.}

\subsection{Related Software}

Although we were not able to find software that scrapes or downloads data from multiple safety databases, we found software that either provides access to a single database or processes data. A large number of software packages are available to either download or process data from the FAERS database. The \texttt{faers} package \citep{faers-bioconductor} accesses the same endpoints used by \texttt{SurVigilance} to download \texttt{FAERS} data. The \texttt{extractFAERS} \citep{extract-faers} and \texttt{faersquarterlydata} \citep{faersquarterlydata} packages also provide various functions to process the data that has been downloaded from the FAERS database. The \texttt{FAERS} package \cite{mlbernau51:online} provides code to set up a \texttt{SQL} database using the \texttt{FAERS} data. The \texttt{OpenVigil} software \citep{OpenVigi18:online} is a set of open-source software packages (\texttt{OpenVigil 1, 2}, and \texttt{OpenVigilFDA}) designed for the analysis of pharmacovigilance data from SRS. It provides tools for data mining, such as disproportionality analysis methods. \texttt{OpenVigil FDA} \citep{bohm2016openvigil} accesses data using the \texttt{openFDA} interface provided by FAERS to access and analyze data. \texttt{OpenVigil 1} and \texttt{2} provide methods to work with data from other sources such as drugbank data. For the VAERS database, three packages \texttt{vaers} \citep{IruckaEm47:online}, \texttt{vaersvax} \citep{vaersvax-cran} and \texttt{vaersNDvax} \citep{vaersNDvax} provide snapshots of the data. The \texttt{vigicaen} package \citep{vigicaen} provides various functionalities to process and analyze data from \texttt{VigiBase}. In Table \ref{tab:comparison}, we present a brief comparison of the packages in terms of various implementation aspects such as programming language, active development, package repository, and coverage. {This list further illustrates the scarcity of software tools that provide access to data from databases other than FAERS and VAERS, alongwith a complete absence of any software that provides access to multiple databases simultaneously, hence underscoring the need for \texttt{SurVigilance}.}

\begin{table}[H]
\centering
\resizebox{1\columnwidth}{!}{%
\begin{tabular}{|l|l|l|l|l|l|l|}
\hline
\rowcolor[HTML]{EEEEEE} 
{\color[HTML]{222222} \textbf{\begin{tabular}[c]{@{}l@{}}Packages/\\ Software\end{tabular}}} &
  {\color[HTML]{222222} \textbf{\begin{tabular}[c]{@{}l@{}}Brief \\ Description\end{tabular}}} &
  {\color[HTML]{222222} \textbf{\begin{tabular}[c]{@{}l@{}}Programming\\ Language\end{tabular}}} &
  {\color[HTML]{222222} \textbf{\begin{tabular}[c]{@{}l@{}}Active\\ Development\end{tabular}}} &
  {\color[HTML]{222222} \textbf{\begin{tabular}[c]{@{}l@{}}Package/Software\\ Repository\end{tabular}}} &
  {\color[HTML]{222222} \textbf{\begin{tabular}[c]{@{}l@{}}Unit tests \\ (Coverage \%)\end{tabular}}} &
  {\color[HTML]{222222} \textbf{\begin{tabular}[c]{@{}l@{}}Continuous\\ Integration\end{tabular}}} \\ \hline

\multicolumn{7}{|l|}{\textbf{FAERS-Focused Tools}} \\ \hline
\textbf{\texttt{faers}}              & Access and analyze FAERS data                               & \texttt{R}               & Yes & \texttt{Bioconductor} &  \textcolor{green}{\cmark} (27.3)  &  \textcolor{green}{\cmark} \\ \hline
\textbf{\texttt{extractFAERS}}       & Process FAERS data                                            & \texttt{R}               & Yes & \texttt{CRAN}         &  \textcolor{red}{\xmark}           &  \textcolor{red}{\xmark} \\ \hline
\textbf{\texttt{faersquarterlydata}} & Process FAERS data                                            & \texttt{R}               & Yes & \texttt{CRAN}         &  \textcolor{green}{\cmark} (99.22) &  \textcolor{red}{\xmark} \\ \hline
\textbf{\texttt{FAERS}}              & Download and pre-process FAERS data                               & Shell Scripting         & No  & \texttt{GitHub}       &  \textcolor{red}{\xmark}           &  \textcolor{red}{\xmark} \\ \hline
\textbf{\texttt{OpenVigil FDA}}      & Access and analyze FAERS data                & \texttt{Java}, \texttt{PHP} & No  & \texttt{SourceForge}  &                                     &  \textcolor{red}{\xmark} \\ \hline

\multicolumn{7}{|l|}{\textbf{VAERS-Focused Tools}} \\ \hline
\textbf{\texttt{vaers}}              & Snapshot of VAERS data                                        & \texttt{R}               & No  & \texttt{GitLab}       &  \textcolor{red}{\xmark}           &  \textcolor{red}{\xmark} \\ \hline
\textbf{\texttt{vaersvax}}           & Snapshot of VAERS data                                        & \texttt{R}               & No  & \texttt{GitLab} and \texttt{CRAN}         &  \textcolor{red}{\xmark}           &  \textcolor{red}{\xmark} \\ \hline
\textbf{\texttt{vaersNDvax}}         & Snapshot of VAERS data                                        & \texttt{R}               & No  & \texttt{GitLab} and \texttt{CRAN}         &  \textcolor{red}{\xmark}           &  \textcolor{red}{\xmark} \\ \hline

\multicolumn{7}{|l|}{\textbf{Other SRS-Focused Tools}} \\ \hline
\textbf{\texttt{vigicaen}}           & Process VigiBase data                                         & \texttt{R}               & Yes & \texttt{CRAN}         &  \textcolor{green}{\cmark} (92.03) &  \textcolor{green}{\cmark} \\ \hline

\multicolumn{7}{|l|}{\textbf{Comprehensive Multi-Database Tools}} \\ \hline
\textbf{\texttt{SurVigilance}}       & \begin{tabular}[c]{@{}l@{}}Access to multiple safety databases. \\Supported databases: FAERS, VAERS, \\VigiAccess, Lareb, Medsafe, DAEN,\\ DMA \end{tabular} & \texttt{Python} & Yes  & \texttt{GitHub} and \texttt{PyPI} &  \textcolor{green}{\cmark} (89.29) &  \textcolor{green}{\cmark} \\ \hline

\end{tabular}%
}
\caption{Comparison of various pharmacovigilance software packages grouped by primary functionality. Active development is determined by the release of any new version between January 1, 2025, and October 14, 2025.}
\label{tab:comparison}
\end{table}

Apart from these packages, the authors in \citep{liu2025mddc} have provided a non-exhaustive list of software which provide methods for signal detection. 

\section{Illustrative examples}

\subsection{Programmatic Data Retrieval from the DMA Database} \label{example:ex-1}

{In this example, we demonstrate the use of functions available in the \texttt{scrapers} module to extract data from the DMA database for the class of alpha-blocker drugs. We specifically consider five drugs: Alfuzosin, Doxazosin, Prazosin, Tamsulosin, and Terazosin. The code to access the data is presented in Listing \ref{code:programmatic-access}. We furthermore create a contingency table with these five drugs, as shown in Table \ref{tab:alpha-blocker-contingency}.}

\begin{lstlisting}[language=Python,caption=\texttt{Python} code demonstrating the use of functions to access data in a programmatic manner without the need for a graphical interface., label=code:programmatic-access]
import os
import pandas as pd

# Import the scraper function to get data from the DMA safety database
from SurVigilance.ui.scrapers import scrape_dma_sb


# List of drugs
DRUGS = ["Alfuzosin", "Doxazosin", "Prazosin", "Tamsulosin", "Terazosin"] 

def get_dma_data():
    out_dir = "dma_out"
    os.makedirs(out_dir, exist_ok=True)
    
    all_results = []
    # Scrape data for each drug and collect results
    
    for drug in DRUGS:
        df = scrape_dma_sb(medicine=drug, output_dir=out_dir, headless=True)
        df = df.copy()
        df["Drug"] = drug
        df["Count"] = pd.to_numeric(df["Count"], errors="coerce").fillna(0)
        all_results.append(df)

        
    combined = pd.concat(all_results, ignore_index=True)

    
    # Build a contingency table
    contingency = combined.pivot_table(
        index="PT", columns="Drug", values="Count", aggfunc="sum", fill_value=0
    )

    
    # Save the contingency table to a CSV file
    contingency_path = os.path.join(out_dir, "dma_contingency_table.csv")
    contingency.to_csv(contingency_path)
    
    return contingency


if __name__ == "__main__":
    get_dma_data()
\end{lstlisting}

\begin{table}[H]
\centering
\resizebox{\columnwidth}{!}{%
\begin{threeparttable}
\begin{tabular}{|l|r|r|r|r|r|}
\hline
\textbf{PT\tnote{a}} &
  \multicolumn{1}{l|}{\textbf{Alfuzosin}} &
  \multicolumn{1}{l|}{\textbf{Doxazosin}} &
  \multicolumn{1}{l|}{\textbf{Prazosin}} &
  \multicolumn{1}{l|}{\textbf{Tamsulosin}} &
  \multicolumn{1}{l|}{\textbf{Terazosin}} \\ \hline
Dizziness & 32 & 9  & 3  & 23 & 2 \\ \hline
Syncope   & 11 & 5  & 7  & 6  & 1 \\ \hline
Fatigue   & 10 & 6  & 8  & 5  & 2 \\ \hline
Headache  & 9  & 10 & 10 & 7  & 1 \\ \hline
\end{tabular}
\begin{tablenotes}
\footnotesize
\item[a] Please note that PT is an abbreviation of Preferred Term. PTs are used to represent specific AEs or symptoms. A detailed definition of PT can be found in \citep{MedDRAHi12:online}.
\end{tablenotes}
\end{threeparttable}
}
\caption{A small portion of the contingency table created using the data for the five alpha-blocker drugs extracted from the DMA database.}
\label{tab:alpha-blocker-contingency}
\end{table}

{Note that the ``Other Drugs'' column is not present in the contingency table presented in Table \ref{tab:alpha-blocker-contingency}. Since databases such as DMA do not allow full database downloads and instead restrict users to only search for specific medical products, a predefined list of drugs is needed to create the ``Other Drugs'' column. The composition of the drugs in that list depends on the kind of analysis the researcher is interested in performing. A more detailed discussion on the construction of the ``Other Drugs'' column is presented in the next example, in Section \ref{sec:construction-of-contingency-table}.}

\subsection{FAERS Data Download and Processing} \label{example:ex-2}
In this section, we present an example that begins with downloading data from the \texttt{FAERS} database using \texttt{SurVigilance}, followed by the construction of a contingency table using the retrieved data.

\subsubsection{Data Download from FAERS}

In order to download the data, the user needs to first instantiate the application as shown in Listing \ref{code:instantiation}.

\begin{lstlisting}[language=Python,caption=\texttt{Python} code for instantiating \texttt{SurVigilance}., label=code:instantiation]
>>> from SurVigilance.ui import UI
>>> UI().run()

  You can now view your Streamlit app in your browser.

  Local URL: http://localhost:8502
  Network URL: http://192.168.12.140:8502
\end{lstlisting}

\begin{figure}[H]
    \centering
    \includegraphics[width=\linewidth]{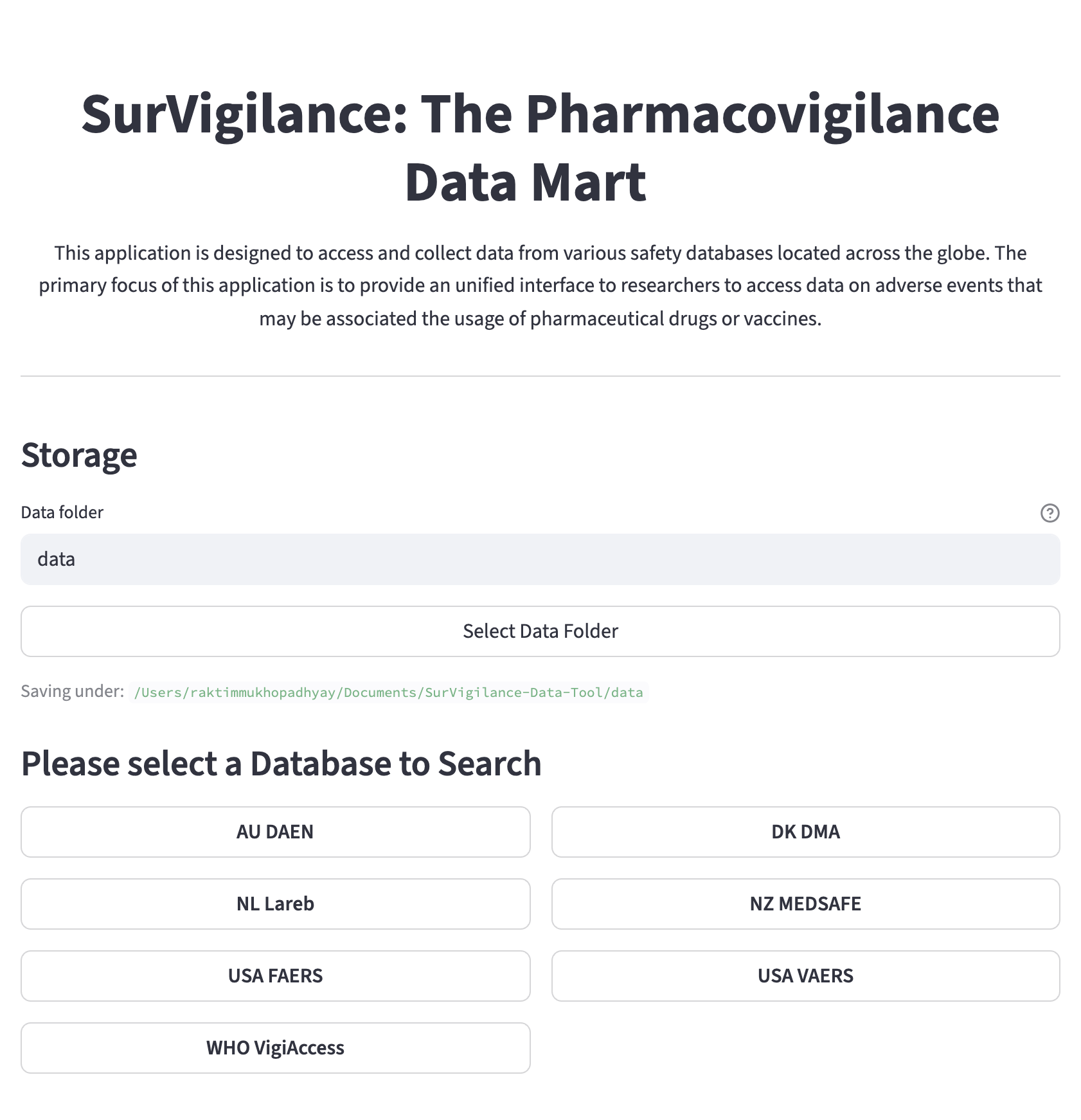}
    \caption{The home page of \texttt{SurVigilance} displaying the available databases. Please note that the home page contains a button at the top which can be used to specify the location where data will be stored.}
    \label{fig:landing}
\end{figure}

\begin{figure}[H]
    \centering
    \includegraphics[width=\linewidth]{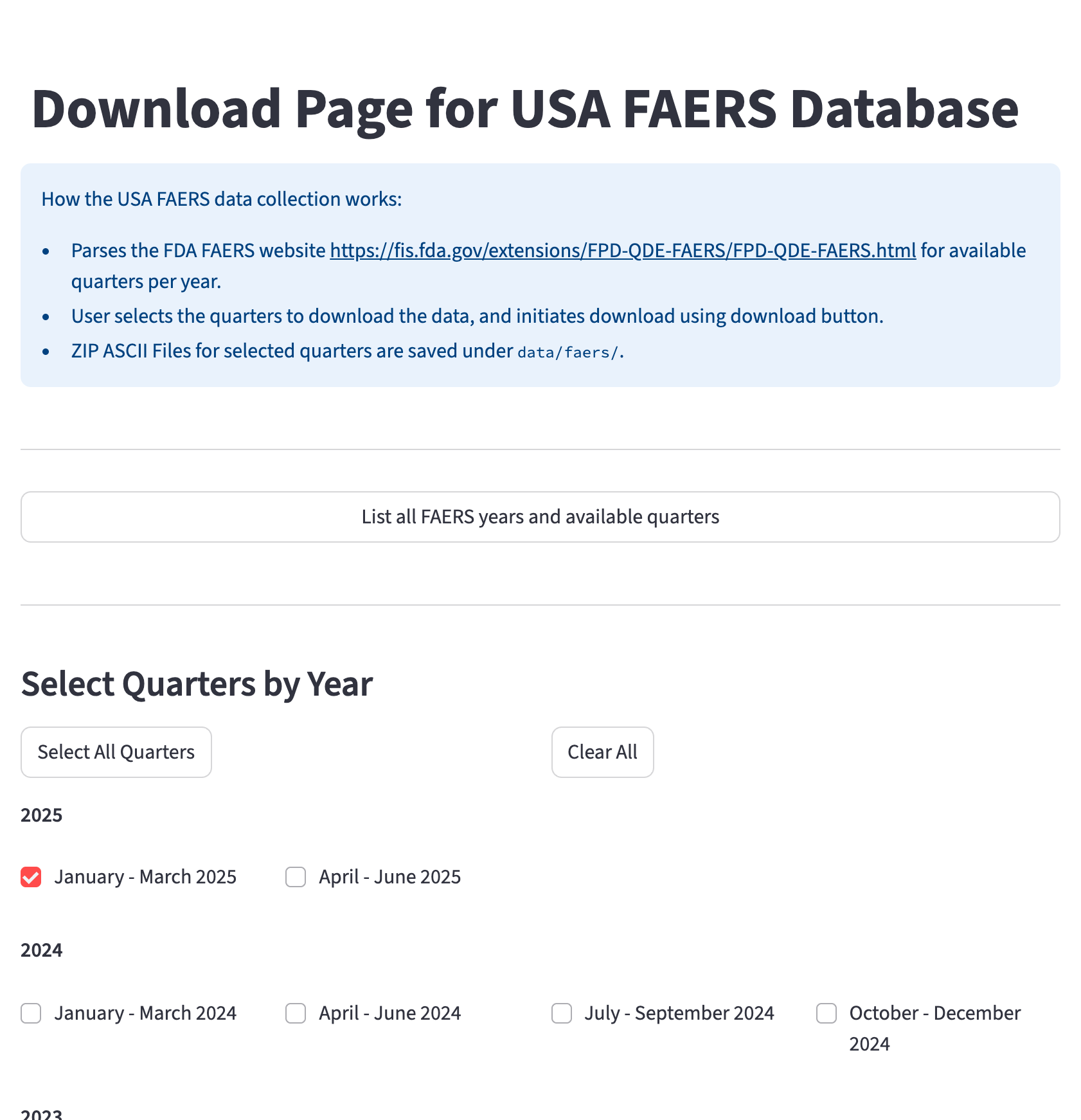}
    \caption{Data download page for FAERS database. After clicking the ``List all FAERS years and available quarters'' button, all available data files are shown from which a user can select the data to be downloaded.}
    \label{fig:search-faers}
\end{figure}

\begin{figure}[H]
    \centering
    \includegraphics[width=\linewidth]{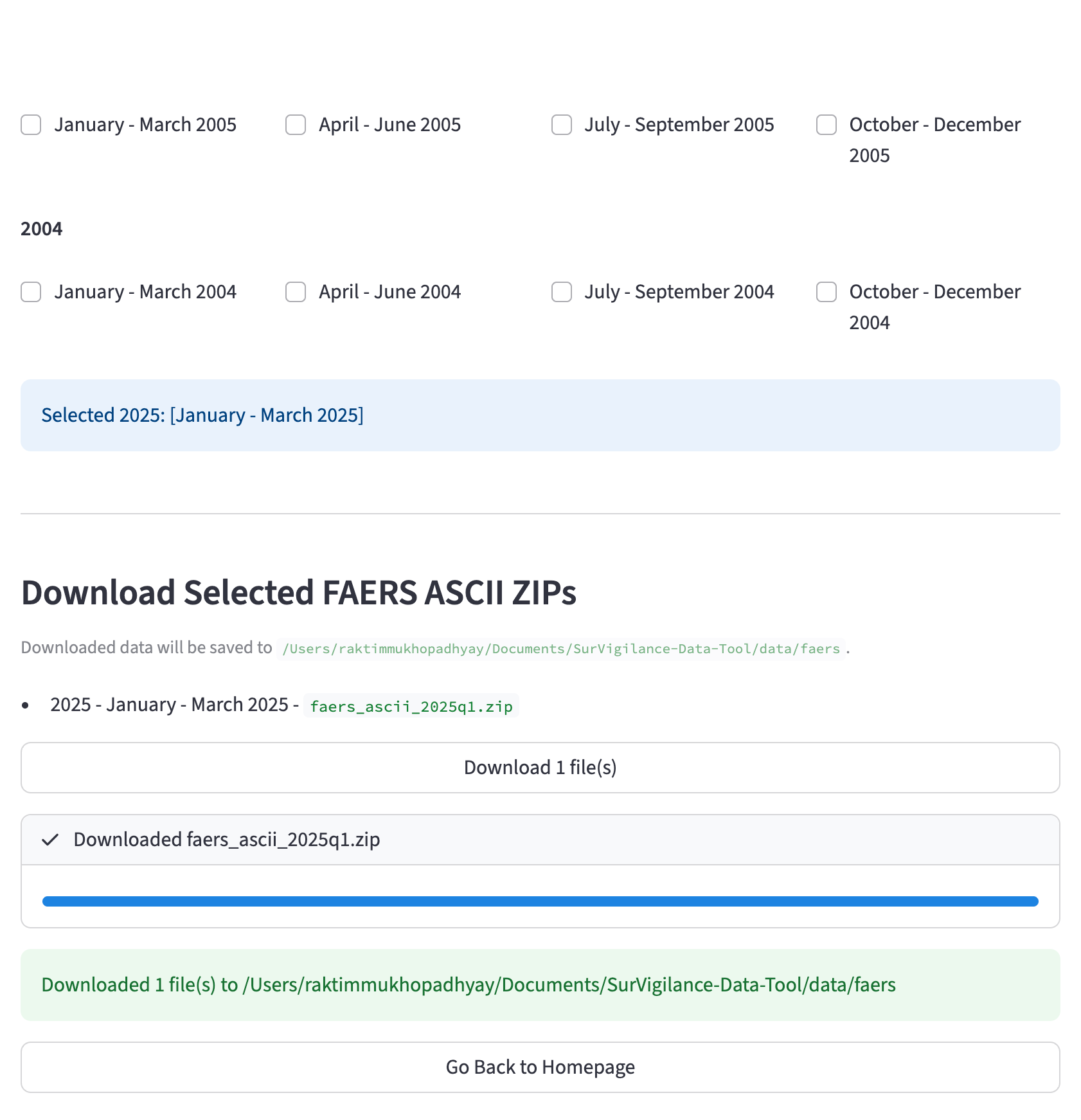}
    \caption{After selecting the data files to be downloaded, the user needs to scroll to the download button at the bottom of the page. Once the download of the selected files is complete, a success message is shown.}
    \label{fig:download}
\end{figure}

This opens the landing page (Figure \ref{fig:landing}). On the landing page, when the user clicks on the \texttt{FAERS} button, the application navigates to the corresponding ``Search Page for \texttt{FAERS} Database'' (Figure \ref{fig:search-faers}). On this page, the user first clicks on the ``List all FAERS years and available quarters'' button, which generates a list of all available quarters for the \texttt{FAERS} database from which data can be downloaded. In this example, we select the January–March 2025 dataset to be downloaded. The user then needs to scroll down and click on the ``Download'' button. Once the download is complete, a completion dialog is displayed (Figure \ref{fig:download}).

\subsubsection{Construction of a Contingency Table from Downloaded FAERS Data} \label{sec:construction-of-contingency-table}

The downloaded data can be used to construct contingency tables suited to a user's needs. In this example, we construct a contingency table of six statin drugs, namely Atorvastatin, Fluvastatin, Lovastatin, Pravastatin, Rosuvastatin, and Simvastatin. A small portion of the constructed contingency table is shown in Table \ref{tab:contingency-table}. The associated code to process the downloaded data to create the contingency table is included in \ref{app:appendix-faers}.

\begin{table}[H]
\centering
\resizebox{\columnwidth}{!}{%
\begin{tabular}{|l|r|r|r|r|r|r|r|}
\hline
\multicolumn{1}{|c|}{\textbf{PT}} &
  \multicolumn{1}{c|}{\textbf{Atorvastatin}} &
  \multicolumn{1}{c|}{\textbf{Fluvastatin}} &
  \multicolumn{1}{c|}{\textbf{Lovastatin}} &
  \multicolumn{1}{c|}{\textbf{Pravastatin}} &
  \multicolumn{1}{c|}{\textbf{Rosuvastatin}} &
  \multicolumn{1}{c|}{\textbf{Simvastatin}} &
  \multicolumn{1}{c|}{\textbf{\begin{tabular}[c]{@{}c@{}}Other \\ Drugs\end{tabular}}} \\ \hline
Myalgia           & 22 & 0 & 1 & 3 & 24 & 17 & 683  \\ \hline
Treatment failure & 2  & 0 & 0 & 3 & 3  & 1  & 1028 \\ \hline
Pain in extremity & 7  & 0 & 0 & 2 & 8  & 0  & 913  \\ \hline
Suicidal ideation & 0  & 0 & 0 & 2 & 0  & 0  & 616  \\ \hline
\end{tabular}%
}
\caption{A small portion of a contingency table for statin drugs constructed using downloaded data from FAERS. Please note for constructing the ``Other Drugs'' column all drugs other than drugs belonging to the statin drug class have been considered. This is in accordance with the drug-based analysis suitable for post-marketing surveillance scenario proposed in \citep{Ding2020-ul}.}
\label{tab:contingency-table}
\end{table}

{The authors in \citep{Ding2020-ul} introduce an interesting dichotomy for constructing contingency tables corresponding to AE-based and drug-based analysis. The AE-based analysis is particularly suited in the context of clinical trials where the objective is to determine whether a specific AE is reported more commonly with a given drug compared to other AEs associated with that same drug. For such analysis, the ``Other Drugs'' column in a $2 \times 2$ contingency table includes all reports associated with other drugs excluding the specific drug of interest. In contrast, the drug-based analysis is suitable for post-marketing surveillance where the interest lies in identifying whether a particular drug is associated with abnormally higher frequencies of AEs than other drugs. In order to avoid confounding from other drugs within the same therapeutic class (e.g., other statin drugs while evaluating atorvastatin), the ``Other Drugs'' column contains the number of reports associated with all drugs outside that therapeutic class. In Table \ref{tab:contingency-table}, the ``Other Drugs'' column contains the number of reports associated with other drugs excluding the statin drug class. This is created keeping in mind drug-based analysis for post-marketing surveillance.}

\section{Computational time associated with accessing data}

{In order to demonstrate the computational efficiency of \texttt{SurVigilance} in retrieving data from various safety databases, we perform a small experiment in which data for ``Atorvastatin'' is retrieved from the five databases, DMA, DAEN, Lareb, Medsafe, and VigiAccess. The data for ``Atorvastatin'' is retrieved 10 times, using the programmatic access functions included in \texttt{SurVigilance} on Python version 3.11.9. The experiment was conducted on an off-the-shelf MacBook Air with an Apple M1 processor running at 3.2 GHz and 16 GB of RAM, using macOS Tahoe. Since the software also needs to access the webpages, internet connectivity is necessary. The experiment was conducted on a 5G Wi-Fi network with a maximum download speed of approximately 450 Mbps. The data retrieval times are shown in Table \ref{tab:time} and the code for running the experiment is included in \ref{app:appendix-retrieval-time}.}

\begin{table}[H]
\centering
\resizebox{0.75\columnwidth}{!}{%
\begin{tabular}{|c|cc|}
\hline
\multirow{2}{*}{\textbf{Database}} & \multicolumn{2}{c|}{\textbf{Data Retrieval Time (in secs)}} \\ \cline{2-3} 
           & \multicolumn{1}{c|}{\textbf{Mean (SD)}} & \textbf{Median [$Q_1$, $Q_3$]} \\ \hline
DMA        & \multicolumn{1}{c|}{29.21 (0.96)}       & 28.98 [28.59, 29.47]     \\ \hline
DAEN       & \multicolumn{1}{c|}{40.35 (1.63)}       & 40.18 [39.70, 41.64]     \\ \hline
Lareb      & \multicolumn{1}{c|}{129.97 (1.15)}      & 129.62 [129.36, 129.68]  \\ \hline
Medsafe    & \multicolumn{1}{c|}{18.40 (0.79)}       & 18.19 [17.81, 18.69]     \\ \hline
VigiAccess & \multicolumn{1}{c|}{57.32 (0.34)}       & 57.25 [57.06, 57.59]     \\ \hline
\end{tabular}%
}
\caption{Computation time associated with retrieving data for Atorvastatin from five safety databases, DMA, DAEN, Lareb, Medsafe, and VigiAccess.}
\label{tab:time}
\end{table}

From Table \ref{tab:time}, we observe that the data retrieval times for the databases are not uniform. In ascending order of data retrieval times (based on the statistics presented in Table \ref{tab:time}), the least time is taken to retrieve data from Medsafe, followed by DMA, DAEN, and VigiAccess, while the maximum time is required for Lareb. The data retrieval times depend on a number of aspects. Medsafe displays the data in a tabular format on its webpage without embedding the data inside collapsible elements. This allows the \texttt{Python} package \texttt{BeautifulSoup} to be used in a straightforward manner to extract the data. On the other hand, the data in DMA is displayed as a table with expandable elements, where all the elements can be expanded with a single button. Since all elements can be expanded using a single button, the overall parsing process remains relatively fast. In the case of DAEN, most of the time is spent waiting for the webpage to prepare the export file. For VigiAccess, the scraper needs to expand each element one by one to extract the underlying data, which increases the retrieval time. In case of the Lareb database, the majority of the retrieval time is spent on loading the search results. Similar to VigiAccess, once the search result is displayed, time is spent on expanding elements and extracting the underlying data. Additionally, while developing the scraping functions for the various databases, we have also embedded intentional delays as described in \ref{app:appendix-responsible}. The delays are embedded into various stages of the data retrieval process. The delay durations are not typically the same across the different steps and databases, as it is also an artifact of a specific database. These delay durations have been set in order to let all elements load in a proper manner in a webpage before progressing to the next step. We have tried to set the delays to an acceptable value which ensures successful scraping and reproducibility in varied conditions (e.g. variable network speed). In certain scenarios, when the delay times were set to lower values, the data retrieval process failed. Hence we have embedded safe delays to maintain the reproducibility of the data retrieval process. 

\section{Limitations}

\texttt{SurVigilance} provides access to data from various safety databases, which primarily provide access to the data through respective webpages. In order for the tool to be able to retrieve data from those webpages, we also need to take into account updates or changes to the structure of a webpage, and implement any additional measures for continued access to the data. \texttt{SurVigilance} might need to be regularly updated to keep it functional, otherwise the tool risks becoming outdated. This challenge exists due to the ephemeral nature of webpages. We have deployed an automated workflow using \texttt{GitHub Actions} that runs the test suite every fortnight. This workflow is designed to fail and trigger notifications to the developer if any of the tests fail, hence allowing us to continuously monitor updates to the database webpages. Additionally, certain databases such as \texttt{VAERS} implement CAPTCHAs, and hence are not completely automatable without human intervention.  

\section{Impact}

\texttt{SurVigilance} aims to streamline the process of accessing pharmacovigilance data. The tool greatly reduces the time and the complexity associated with accessing data from safety databases across the world. By providing both graphical and programmatic interfaces to the underlying methods, \texttt{SurVigilance} is appropriately suited for its integration in pre-existing analytical pipelines. \texttt{SurVigilance} has been designed keeping in mind the maintainability and extensibility of the codebase in the long term, supporting additional databases in the future. The tool furthermore gives users access to the data in raw form, without any additional processing that researchers might find unnecessary or unsuited to their needs. A search on Google Scholar for the term ``pharmacovigilance'' returns approximately 25,000 search results since 2021, highlighting the extent of the research in the domain. We hypothesize the tool will be of practical importance to researchers in this domain. \texttt{SurVigilance} thus addresses a key methodological barrier in the domain by streamlining data acquisition while leaving analytical choices to the investigator.

\section{Conclusions}

{In this work, we introduced \texttt{SurVigilance}, a modular, open-source tool that simplifies access to pharmacovigilance data from multiple global safety databases. A comparison with other similar software packages is included in Table \ref{tab:comparison}, which illustrates that most software solutions focus on a single database (FAERS) while other packages focus on providing helper functions for data processing from FAERS or VigiAccess. A number of packages also provide snapshots of the VAERS database. We have also demonstrated how our tool can be used to download data using the graphical interface and subsequently processed to create contingency tables used in various methods in pharmacovigilance research. In addition to the graphical interface, \texttt{SurVigilance} also offers programmatic access to the databases for more advanced users, with reasonable data retrieval times. While the tool currently supports seven databases, the modular structure of the codebase allows for further extension of the software in the future. In conclusion, \texttt{SurVigilance} aims to further global pharmacovigilance research by removing barriers to access safety data. }

\section*{CRediT authorship contribution statement}

{\textbf{Raktim Mukhopadhyay:} Writing - original draft, Writing - review \& editing, Validation, Software, Investigation.
\textbf{Marianthi Markatou:} Conceptualization, Validation, Investigation, Resources, Writing - original draft, Writing - review \& editing, Supervision, Project administration, Funding acquisition.}

\section*{Declaration of competing interest}

The authors declare that they have no known competing financial interests or personal relationships that could have appeared to influence the work reported in this paper.

\section*{Acknowledgments}

We gratefully acknowledge the organizations that maintain the various safety databases included in our work, whose efforts make these datasets publicly accessible. The senior author acknowledges financial support in the form of a research award from KALEIDA Health Foundation, USA (Grant Number 82114), that supported the work of the first author.

\section*{Data availability}

All data shown in this work are publicly available from the respective safety databases. The data for Example \ref{example:ex-1} can be accessed from \url{https://laegemiddelstyrelsen.dk/sideeffects/side-effects-of-medicines/interactive-adverse-drug-reaction-overviews/.aspx} and the data for Example \ref{example:ex-2} can be found at \url{https://fis.fda.gov/extensions/FPD-QDE-FAERS/FPD-QDE-FAERS.html}. The associated \texttt{Python} scripts to download and process the data are also included as part of this work.

\appendix

\section{Responsible Use}\label{app:appendix-responsible}

We extensively referred to the Responsible Use guidance provided as part of the publication  \citep{TRUONG2025102373}, which is available at \url{https://github.com/truongbavinh/HTMLDownloader-/blob/master/docs/LEGAL.md}. 

We have undertaken the following measures to promote responsible use:

\begin{enumerate}
    \item Adhering to the guidelines mentioned above, we have not implemented any concurrency or parallelism to prevent overwhelming the servers with traffic for retrieving the data.
    \item We have embedded delays while scraping data to further prevent stressing the server and emulate human-like access patterns. This can be clearly seen from the data retrieval times illustrated in Table \ref{tab:time}.
    \item We did not attempt to circumvent additional security measures such as CAPTCHAs which is present for accessing the VAERS database.
    \item We verify the presence of \texttt{robots.txt} files and make sure that the scraping of the database is not prohibited. The URLs to access the \texttt{robots.txt} for the various databases is shown in Table \ref{tab:robots}.
    \item The terms of use associated with the various webpages to access the databases can be found in Table \ref{tab:terms}. To the best of our knowledge, these webpages do not explicitly prohibit web scraping. 
\end{enumerate}

\begin{table}[H]
    \centering
    \begin{tabular}{|l|l|l|}
    \hline
        \textbf{Database} & \textbf{\texttt{robots.txt} Present} & \textbf{Link} \\ \hline
        DAEN & Yes & \url{https://www.tga.gov.au/robots.txt} \\ \hline
        DMA & No & NA \\ \hline
        FAERS & Yes &  \url{https://www.fda.gov/robots.txt} \\ \hline
        Lareb & No & NA \\ \hline
        Medsafe & Yes & \url{https://www.medsafe.govt.nz/robots.txt} \\ \hline
        VAERS & No & NA \\ \hline
        VigiAccess & No & NA \\ \hline
    \end{tabular}
    \caption{URLs associated with the \texttt{robots.txt} files for various databases.}
    \label{tab:robots}
\end{table}

\begin{table}[H]
\centering
\begin{tabular}{|l|p{11cm}|}
\hline
\textbf{Database} & \textbf{Link for Terms of Use} \\ \hline
DAEN & \url{https://www.tga.gov.au/about-tga/using-our-website/copyright} \\ \hline
DMA & \url{https://laegemiddelstyrelsen.dk/en/sideeffects/side-effects-of-medicines/interactive-adverse-drug-reaction-overviews/terms-of-use-of-the-interactive-adr-overviews/} \\ \hline
FAERS & \url{https://www.fda.gov/about-fda/about-website/website-policies} \\ \hline
Lareb & \url{https://www.lareb.nl/en/pages/privacy-statement/} \\ \hline
Medsafe & \url{https://www.medsafe.govt.nz/other/siteinfo.asp} \\ \hline
VAERS & \url{https://vaers.hhs.gov/privacy.html} \\ \hline
VigiAccess & \url{https://www.vigiaccess.org/} \\ \hline
\end{tabular}
\caption{URLs for the ``Terms of Use/Service'' associated with the webpages used to access the databases.}
\label{tab:terms}
\end{table}

\section{Code for Processing Downloaded FAERS Data}
\label{app:appendix-faers}

\begin{lstlisting}[language=Python,caption=\texttt{Python} code for processing the downloaded FAERS data to generate a contingency table for statin drugs., label=code:appendix-a-code]
import pandas as pd
from pathlib import Path
import zipfile

# Unzip once into a local folder
zip_path = Path(
    "data/faers/faers_ascii_2025q1.zip"
)
out_dir = Path("extracted_files")
out_dir.mkdir(parents=True, exist_ok=True)
with zipfile.ZipFile(zip_path, "r") as zf:
    zf.extractall(out_dir)

# Get the paths for the DRUG* and REAC* files
drug_path = next(out_dir.rglob("ASCII/DRUG*.txt"))
reac_path = next(out_dir.rglob("ASCII/REAC*.txt"))

# Read both files ($ delimited)
read_options = dict(sep=r"\$", engine="python", dtype=str)
drug = pd.read_csv(drug_path, **read_options)
reac = pd.read_csv(reac_path, **read_options)

# Keep just the columns we care about from DRUG and filter to PS/SS
drug = drug.rename(columns=str.lower)[
    ["primaryid", "caseid", "drug_seq", "role_cod", "prod_ai"]
]
reac = reac.rename(columns=str.lower)
drug = drug[drug["role_cod"].isin(["PS", "SS"])]

# Join DRUG and REAC on primaryid + caseid
dat = pd.merge(drug, reac, on=["primaryid", "caseid"], how="outer")

# For each caseid, keep the row with the largest primaryid (treat that as "most recent")
dat["_pid_num"] = pd.to_numeric(dat["primaryid"], errors="coerce")
dat = (
    dat.sort_values(["caseid", "_pid_num"])
    .groupby("caseid", as_index=False)
    .tail(1)
    .drop(columns="_pid_num")
)


# Make statins in ALL CAPS; everything else becomes OTHER
def label_statin(s):
    s = (s or "").lower()
    for d in [
        "atorvastatin",
        "fluvastatin",
        "lovastatin",
        "pravastatin",
        "rosuvastatin",
        "simvastatin",
    ]:
        if d in s:
            return d.upper()
    return "OTHER"


dat["DRUG_NEW"] = dat["prod_ai"].apply(label_statin)

# Keep only PTs that actually co-occur with a statin
statin_pts = dat.loc[dat["DRUG_NEW"] != "OTHER", "pt"].dropna().unique()
dat = dat[dat["pt"].isin(statin_pts)]

# Build the contingency table
pt_table = (
    pd.pivot_table(
        dat,
        index="pt",
        columns="DRUG_NEW",
        values="primaryid",
        aggfunc="count",
        fill_value=0,
    )
    .reset_index()
    .rename(columns={"pt": "PT"})
)
pt_table.columns = [c.upper() for c in pt_table.columns]
pt_table = pt_table[
    [c for c in pt_table.columns if c != "OTHER"]
    + (["OTHER"] if "OTHER" in pt_table.columns else [])
]

# Save the file
pt_table.to_excel("STATIN_ALL_PT.xlsx", index=False)
\end{lstlisting}

\section{Code to Measure Data Retrieval Time for Atorvastatin}
\label{app:appendix-retrieval-time}

\begin{lstlisting}[language=Python,caption=\texttt{Python} code to record data retrieval times for Atorvastatin from multiple safety databases., label=code:appendix-a-code]
import sys
import time
import pandas as pd

# settings
RUNS = 10
DRUG = "atorvastatin"
HEADLESS = True

from SurVigilance.ui.scrapers import (
    scrape_daen_sb,
    scrape_dma_sb,
    scrape_lareb_sb,
    scrape_medsafe_sb,
    scrape_vigiaccess_sb,
)

def time_call(fn):
    start = time.perf_counter()
    fn()
    return time.perf_counter() - start

def main():
    targets = {
        "DMA": lambda: scrape_dma_sb(medicine=DRUG, headless=HEADLESS),
        "DAEN": lambda: scrape_daen_sb(medicine=DRUG, headless=HEADLESS),
        "Lareb": lambda: scrape_lareb_sb(medicine=DRUG, headless=HEADLESS),
        "Medsafe": lambda: scrape_medsafe_sb(
            searching_for="medicine", drug_vaccine=DRUG, headless=HEADLESS
        ),
        "VigiAccess": lambda: scrape_vigiaccess_sb(medicine=DRUG, headless=HEADLESS),
    }

    print(f"Recording time for '{DRUG}' runs per database={RUNS}\n")

    results = {name: [] for name in targets}

    # round-robin execution
    for r in range(1, RUNS + 1):
        for name, fn in targets.items():
            dur = time_call(fn)
            results[name].append(dur)
            print(f"  {name:10s} run {r:02d}: {dur:.2f}s")

    print("\n Descriptive Summary")

    summary_data = []
    for name in targets:
        s = pd.Series(results[name])
        mean = s.mean()
        sd = s.std()
        median = s.median()
        q1 = s.quantile(0.25)
        q3 = s.quantile(0.75)

        summary_data.append({
            "Database": name,
            "Mean (SD)": f"{mean:.2f} ({sd:.2f})",
            "Median [Q1, Q3]": f"{median:.2f} [{q1:.2f}, {q3:.2f}]"
        })

        print(
            f"{name}  n={len(s)}  mean={mean:.2f}s  median={median:.2f}s  sd={sd:.2f}s"
        )

    df_summary = pd.DataFrame(summary_data)
    output_file = f"scraper_timing_summary_{DRUG}.csv"
    df_summary.to_csv(output_file, index=False)

if __name__ == "__main__":
    main()
\end{lstlisting}

 \bibliographystyle{elsarticle-num} 
 \bibliography{ref}








\end{document}